\documentclass{article}
\usepackage{spconf,amsmath,graphicx}

\def\evec{\mathbf e}

\def\xvec{\mathbf x}

\def\muvec{\boldsymbol \mu}

\usepackage{lineno,hyperref}
\usepackage{cite}
\usepackage{amsmath,graphicx}
\usepackage{float}
\usepackage{amssymb,amsmath,comment,caption, subcaption, graphicx,amsmath,mathrsfs, array, hyperref, multirow}
\usepackage{booktabs}
\usepackage{diagbox}
\usepackage{xcolor}
\modulolinenumbers[5]
\ninept

\usepackage[square,sort,comma,numbers]{natbib}
\title{x-vectors meet emotions:  A study on dependencies between  \\
emotion and speaker recognition}
\name{Raghavendra Pappagari\sthanks{Both the authors contributed equally to this paper}, Tianzi Wang\footnotemark[1], Jes\'us Villalba, Nanxin Chen, Najim Dehak}
\address{Center for Language and Speech Processing \\
    Johns Hopkins University, Baltimore, MD \\
    \{rpappag1, wtianzi1, jvillal7, bobchennan, ndehak3\}@jhu.edu
    }

\begin{document}
%
\maketitle
\begin{abstract}
In this work, we explore the dependencies between speaker recognition and emotion recognition. 
We first show that knowledge learned for speaker recognition can be reused for emotion recognition through transfer learning.
Then, we show the effect of emotion on speaker recognition.
For emotion recognition, we show that using a simple linear model is enough to obtain good performance on the features extracted from pre-trained models such as the x-vector model.
Then, we improve emotion recognition performance by fine-tuning for emotion classification.
We evaluated our experiments on three different types of datasets: IEMOCAP, MSP-Podcast, and Crema-D. 
By fine-tuning, we obtained 30.40\%, 7.99\%, and 8.61\% absolute improvement on IEMOCAP, MSP-Podcast, and Crema-D respectively over baseline model with no pre-training.
Finally, we present results on the effect of emotion on speaker verification.
We observed that speaker verification performance is prone to changes in test speaker emotions.
We found that trials with angry utterances performed worst in all three datasets. 
We hope our analysis will initiate a new line of research in the speaker recognition community.

\end{abstract}
\section{Introduction}
\label{sec:intro}

In this work, we explore the dependencies between speaker recognition and speech emotion recognition (SER).
Speaker verification, a more general task of speaker recognition, deals with verifying speaker identity in a pair of utterances. 
The goal of the SER task is to recognize the emotional state of a speaker in a speech recording.
For both tasks, acoustic parameters such as pitch, fundamental frequencies, acoustic energy play a crucial role in obtaining better performance.
Hence, we hypothesize that models trained to discriminate speakers can be reused for SER.

Some of the applications of speaker verification include voice-based authentication, security systems, and personal assistants. 
SER is useful in applications such as detecting hate speech in social media, detecting patient's emotions, call routing based on emotion, actors analysis in the entertainment industry, mental health analysis, and human-machine interaction.


Several works in the past have tried to improve SER by using various feature representations and models.
In~\cite{tzirakis2017end, sarma2018emotion, trigeorgis2016adieu} feature learning from raw-waveform or spectrogram using CNN, LSTM based models is explored.
In~\cite{cho2018deep, zhao2019speech, huang2014speech, lim2016speech}, CNN and LSTM based models are explored from feature representations such as MFCC and OpenSMILE~\cite{eyben2013recent} features.
In~\cite{latif2018adversarial, han2018towards, parthasarathy2019improving, sahu2018enhancing}, adversarial learning paradigm is explored for robust recognition.
In~\cite{latif2018transfer, lakomkin2018reusing}, transfer learning approach is explored.

In this work, we follow the transfer learning approach.
Our work is motivated by several previous works~\cite{lakomkin2018reusing, raj2019probing, williams2019disentangling}.
It is shown in~\cite{lakomkin2018reusing} that reusing an ASR model trained to predict phonemes is helpful for the SER task.
In~\cite{raj2019probing}, authors studied the applicability of speaker based utterance representations such as i-vectors and x-vectors for several downstream tasks related to speech, speaker, and utterance meta information. 
However, they did not study for emotion-related tasks.

Authors in~\cite{williams2019disentangling} show that speaker-based utterance-level representations i-vectors and x-vectors encode speaking-style information and emotion.
However, their experimental setup included overlapping speakers between training and testing data splits. We believe that speaker overlap should be avoided in SER tasks, especially when using speaker-specific representations as input.
In this paper, we present results using pre-trained as well as fine-tuned models which is not studied in~\cite{williams2019disentangling}.


In this paper, we explore transfer learning for SER task from neural networks trained to discriminate speakers such as the x-vector model.
First, we show that emotion-related information is encoded in x-vectors, and then we show that fine-tuning for emotion targets further improves the performance.
We use two pre-trained models for this study--one trained with augmentation and another without augmentation.
We also experiment with augmenting the emotion data for better performance.
Then, we present results of speaker verification on emotion datasets and show the effect of emotion on its performance, which could potentially initiate a new line of research in the speaker recognition community.

The main contributions of this work are:
\begin{itemize}
    \item Exploring pre-trained models trained to discriminate speakers for emotion tasks on 3 different types of datasets
    \item Fine-tuned models for SER task
    \item Results with data augmentation on emotion datasets
    \item Analysis of the effect of emotion on speaker verification results
\end{itemize}

Rest of the paper is organized as follows. In Section~\ref{sec:our_approach}, we present our method followed by experimental setup in Section~\ref{sec:expt_setup}. Then, we discuss results in Section~\ref{sec:results} and finally in Section~\ref{sec:conclusion}, we present conclusion and future work.

\section{Our Approach}
\label{sec:our_approach}

In this section, we present details of the x-vector model reused for the SER task.
Then, we explain the transfer learning approach followed to perform SER.
It is shown in the literature that i-vectors and x-vectors perform well on speaker related tasks such as speaker verification~\cite{villalba2019state}, speaker diarization \cite{shum2013unsupervised,sell2014speaker,maciejewski2018characterizing,sell2018diarization}.
In this work, we only exploit the x-vector model because of its superiority over i-vectors~\cite{snyder2018x} and also because it is easy to adapt for down-stream tasks.

\subsection{x-Vector Model}
In this paper, we used state-of-the-art ResNet x-vector model reported in~\cite{villalba2019state} for utterance level speaker embedding extraction.
The network consisted of three parts: frame-level representation learning network, pooling network, and utterance-level classifier.
Frame-level representation learning network uses ResNet-34~\cite{he2016deep} structure, which consists of several 2D convolutional layers with short-cut connections between them.
After that, we used a multi-head attention layer to summarize the whole utterance into a large embedding.
This layer takes ResNet outputs $\xvec_t$ as input and computes its own attention scores ${w_{h,t}}$ for each head $h$:
\begin{align}
  w_{h, t} = \frac{\exp(-s_h \left\|\xvec_t-\muvec_h\right\|)}{\sum_{t=1}^T \exp(-s_h \left\|\xvec_t-\muvec_h\right\|)} \;.
\end{align}
Attention scores $w_{h, t}$ are normalized along time axis.

Output embedding for head $h$ is the weighted average over its inputs:
\begin{align}
    \evec_h = \sum_t w_{h,t} \xvec_t
\end{align}
Different heads are designed to capture different aspects of input signal.
Embedding from different heads are concatenated and projected by an affine transformation into the final embedding. From the pooling layer to output, there are two fully connected layers, and it predicts speaker identity in the training set.
Angular softmax~\cite{liu2017sphereface} loss was used to train the network.
The whole network structure is illustrated in Table~\ref{tab:xvec_arch}.
For more details, please refer to~\cite{villalba2019state}.

\begin{table}
\centering
\begin{tabular}{|c|c|c|}
\hline
Component  & Layer   & Output Size   \\ \hline
\multirow{6}{*}{\begin{tabular}[c]{@{}l@{}}Frame-level\\Representation\\ Learning\end{tabular}} & $7 \times 7, 16$                                                                          & $T \times 23$        \\ \cline{2-3} 
& \begin{tabular}[c]{@{}l@{}} $\begin{bmatrix} 3 \times 3, 16 \\ 3 \times 3, 16 \end{bmatrix} \times 3$ \end{tabular}             & $T \times 23$        \\ \cline{2-3} 
& \begin{tabular}[c]{@{}l@{}} $\begin{bmatrix} 3 \times 3, 32 \\ 3 \times 3, 32 \end{bmatrix} \times 4$, stride 2\end{tabular}   & $\frac{T}{2} \times 12$    \\ \cline{2-3} 
& \begin{tabular}[c]{@{}l@{}} $\begin{bmatrix} 3 \times 3, 64 \\ 3 \times 3, 64 \end{bmatrix} \times 6$, stride 2\end{tabular}   & $\frac{T}{4} \times 6$     \\ \cline{2-3} 
& \begin{tabular}[c]{@{}l@{}} $\begin{bmatrix} 3 \times 3, 128 \\ 3 \times 3, 128 \end{bmatrix} \times 3$, stride 2\end{tabular} & $\frac{T}{8} \times 3$     \\ \cline{2-3} 
& average pool $1 \times 3$                                                            & $\frac{T}{8}$         \\ \hline
Pooling                                                                                       & 32 heads attention                                                               & $32 \times 128$      \\ \hline
\multirow{2}{*}{\begin{tabular}[c]{@{}l@{}}Utterance-level\\ Classifier\end{tabular}}         & FC                                                                               & 400           \\ \cline{2-3} 
                                                                                              & FC                                                                               & \#spk:12,872 \\ \hline
\end{tabular}
\caption{ResNet architecture used in the x-vector model}
\label{tab:xvec_arch}
\end{table}

\subsection{Emotion Recognition}
\label{subsec:emotion_recog}

From a pre-trained x-vector model, we can transfer knowledge to achieve SER in two ways:
\begin{itemize}
    \item Extract x-vectors and apply a simple linear model like logistic regression (LR)
    \item Replace the speaker-discriminative output layer with emotion-discriminative layer and fine-tune
\end{itemize}

In this paper, we show experiments with both methods. 
We compare these two methods with widely used OpenSMILE features.
We also experiment with two versions of pre-trained x-vector models: one trained with augmentation, referred to as \textit{ResNet-aug}, and another trained with only clean data, referred to as \textit{ResNet-clean}.

\section{Experimental Setup}
\label{sec:expt_setup}

\begin{table*}
    \centering
    \begin{tabular}{@{}p{6cm}|c|c|c|c|c|c@{}}
      \toprule
       Dataset Name & \multicolumn{2}{c|}{IEMOCAP} & \multicolumn{2}{c|}{MSP-Podcast} & \multicolumn{2}{c}{Crema-D} \\
        \midrule
      Emotion Classification Training data   & \textit{Clean}& \textit{Clean+aug}  & \textit{Clean}& \textit{Clean+aug}  & \textit{Clean}& \textit{Clean+aug}  \\
    \midrule
    Randomly initialized ResNet (GeMAPS) & 45.14 & 49.42 & 51.36 & 51.42 & 74.44 & 74.39  \\
    Randomly initialized ResNet (MFCC) & 39.58 & 48.23 & 50.47 & 48.87 & 72.93 & 75.20  \\
    \midrule
    \midrule
    Pre-trained \textit{ResNet-clean} (MFCC) -- Frozen & 59.05  &  54.56  & 56.75 & 57.10 & 79.03 & 78.86  \\
    Pre-trained \textit{ResNet-aug} (MFCC) -- Frozen & 56.11 & 56.44 & 52.58 & 54.59  & 75.65 & 77.49    \\
    \midrule
    \midrule
    Fine-tuned \textit{ResNet-clean} (MFCC) &  65.95 & 59.15  &57.42  & 57.07 & 76.00 & 80.00  \\
    Fine-tuned \textit{ResNet-aug} (MFCC)& 60.25  & \textbf{70.30}  & \textbf{58.46} & 56.70   & 80.55 & \textbf{81.54}  \\
    
    \bottomrule
    \end{tabular}
    \caption{SER results on three datasets. In the first column, \textit{ResNet-clean} and \textit{ResNet-aug} denotes unaugmented and augmented x-vector models. Text in the paranthesis denotes the feature set we used to train.
    In the second row, \textit{Clean} denotes emotion classification training is only on clean data and \textit{Clean+aug} denotes clean data is augmented with noisy data for the respective datasets. All the numbers in this table are weighted f-scores for the respective datasets.}
    \label{tab:all_emotion_results}
\end{table*}

\subsection{Datasets}
We validate our experiments on three different types of datasets: IEMOCAP (acted and no restriction on spoken content), MSP-Podcast (natural and no restriction on spoken content), and Crema-D (acted and restricted to 12 sentences). The details of each dataset are as follows.

\subsubsection{IEMOCAP}
IEMOCAP dataset is a multimodal dyadic conversational dataset recorded with 5 female and 5 male actors~\cite{busso2008iemocap}.
It contains conversations from 5 sessions wherein each session one male and female actor converse about a pre-defined topic.
Each session is segmented into utterances manually, and each utterance is annotated by at least 3 annotators to categorize into one of 8 emotion classes (angry, happy, neutral, sad, disgust, fear, excited).
Conversations are scripted and improvisational in nature.


In this work, we followed previous works in choosing data for our experiments.
We combined happy and excited emotions into one class. 
We choose a subset of data consisting of 4 emotions: angry, sad, neutral, happy.
As the number of speakers and utterances in this dataset is low, we opted for 5-fold cross-validation (CV) to obtain reliable results.
As it was shown in~\cite{raj2019probing} that speaker verification models capture session variability along with speaker characteristics; we did leave-one-session-out training for 5-fold CV to avoid overlapping of speakers and sessions between training and testing.
In each fold, we used weighted f-score as our metric, and hence, we reported an average of weighted f-scores of 5-fold CV for each experiment.

\subsubsection{MSP-Podcast Dataset}
MSP-Podcast corpus\footnote{Data provided by The University of Texas at Dallas through the Multimodal Signal Processing Lab. This material is based upon work supported by the National Science Foundation under Grants No. IIS-1453781 and CNS-1823166. Any opinions, findings, and conclusions or recommendations expressed in this material are those of the author(s) and do not necessarily reflect the views of the National Science Foundation or the University of Texas at Dallas.} is collected from podcast recordings. 
The recordings were processed to remove segments with SNR less than 20dB, background music, telephone quality speech, and overlapping speech. For more information on this dataset, please refer to~\cite{lotfian2017building}.
In this work, we used 5 emotions: angry, happy, sad, neutral, disgust for classification as in~\cite{lotfian2019curriculum}. 
We used the standard splits in Release 1.4 for training, development, and testing.
This dataset has 610 speakers in the training split, 30 in the development, and 50 speakers in the test split.

\subsubsection{Crema-D Dataset}
Crema-D dataset\footnote{https://github.com/CheyneyComputerScience/CREMA-D} is a multimodal dataset (audio and visual) with 91 professional actors enacting a target emotion for a pre-defined list of 12 sentences.
It includes 48 male and 48 female actors with a diverse ethnicity and age distribution.
In this work, we use 4 emotion categories: angry, happy, sad, neutral discarding disgust, and fear to balance the dataset.
We used 51 actors in training, 8 for development, and 32 for testing.

\subsection{Feature Extraction}
In this work, we extracted 23-dim MFCC with a 10ms frame shift and 25ms frame size. We used a simple energy-based speech activity detector to remove silence segments from the utterances.
Our pre-trained models were trained with MFCC.
For OpenSMILE features, referred to as GeMAPS, we extracted 88-dim features as suggested in~\cite{eyben2015geneva} with a 10ms frame shift and 25ms frame size.

\section{Results}
\label{sec:results}


\subsection{Emotion recognition}


Table~\ref{tab:all_emotion_results} presents results of SER task with ResNet architecture on all three datasets. 
As noted in Section~\ref{subsec:emotion_recog}, \textit{ResNet-clean} and \textit{ResNet-aug} denotes unaugmented and augmented x-vector models.
In the second row, \textit{Clean} denotes emotion classification training is only on clean data, and \textit{Clean+aug} denotes clean data is augmented with noisy data for the respective datasets. 
Augmentation is done with additive noise and music using MUSAN corpus~\cite{snyder2015musan}.

Comparison of 3rd and 4th rows suggests GeMAPS perform better than MFCC in most cases, but as our pre-trained models were trained with MFCC, we did not consider GeMAPS for further experiments.
Significant improvements were obtained on all the datasets by using pre-trained models compared to random initialization suggesting that pre-training is helpful.
In \textit{Clean} setting i.e., when using only clean data for emotion classification, pre-trained \textit{ResNet-clean} performed 2.94\%, 4.17\% and 3.38\% better than pre-trained \textit{ResNet-aug} on IEMOCAP, MSP-Podcast and Crema-D respectively. 
A similar conclusion was reported in~\cite{raj2019probing} for tasks such as prediction of the session, utterance length, gender, etc..
Having observed the good performance of pre-trained ResNet models, which are trained to discriminate speakers, we proceeded to fine-tune the pre-trained models for emotion recognition.
By fine-tuning, we obtained improvements in all cases except for a 3.03\% drop on Crema-D when using \textit{ResNet-clean}.

From comparison of \textit{Clean+aug} column with \textit{Clean}, we can observe that augmenting data for emotion classification helped on IEMOCAP and Crema-D datasets except when using \textit{ResNet-clean} suggesting that it is not robust to noise.
For the MSP-Podcast dataset, improvements are not clear using data augmentation. 
We obtained improvements with fixed pre-trained models but not when fine-tuning for emotion task.

Overall, fine-tuned \textit{ResNet-aug} model worked best on IEMOCAP and Crema-D in \textit{Clean+aug} setting with 70.30\% and 81.54\% respectively.
On MSP-Podcast dataset, fine-tuned \textit{ResNet-aug} worked best with 58.46\% on clean training data.
In terms of absolute improvement over randomly initialized ResNet (MFCC) baseline, we obtained 30.40\%, 7.99\%, and 8.61\% on IEMOCAP, MSP-Podcast, and Crema-D, respectively.
It is difficult to compare our results with previous works as there are no standard splits for IEMOCAP and Crema-D. 
In the case of MSP-Podcast, the dataset collection is an ongoing effort, and we did not find previous works on the current release yet.

\begin{figure}
    \centering
    \includegraphics[width=0.5\textwidth]{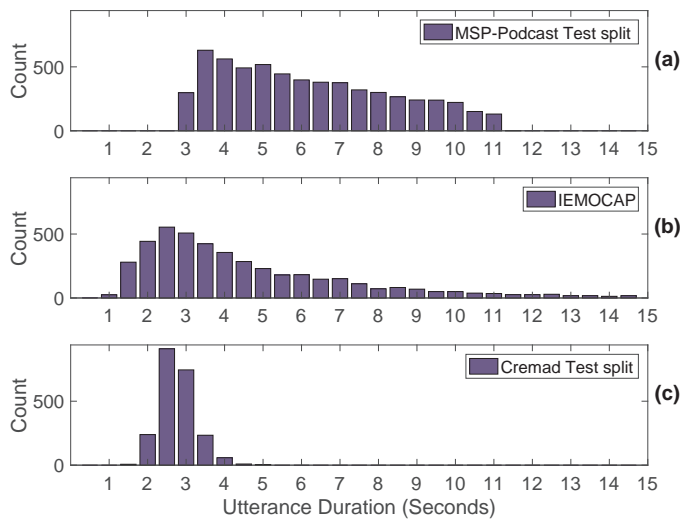}
   \caption{Histograms of utterance durations}
 \label{fig:duration_hist}
\end{figure}

\subsection{Speaker Verification}

In this section, we show the effect of emotion on the performance of the speaker verification system. 
For this experiment, we have formed speaker verification trials by comparing every utterance against each other.  Thus, we obtained cross-emotion and same-emotion trials.
We did not consider cross-gender trials as they are relatively easier than same-gender trials. 
Table~\ref{tab:eer_iemocap},~\ref{tab:eer_msp} and~\ref{tab:eer_cremad} presents speaker verification results in terms of EER for IEMOCAP, MSP-Podcast and Crema-D datasets respectively. The results are isolated given the emotion of the enrollment (rows) and test utterances (columns). 

EERs on IEMOCAP are very high because utterances are very short and because of domain mismatch.
Histogram of IEMOCAP dataset utterance duration is presented in Fig.~\ref{fig:duration_hist}(b).
The majority of the utterances in the dataset are less than 4 seconds.
EERs for MSP-Podcast dataset are better than IEMOCAP but still above 10\%, which can be attributed to the short utterances in the dataset.
Histogram of MSP-Podcast dataset utterance duration is presented in Fig.~\ref{fig:duration_hist}(a). 
It can be observed that most utterances are short but longer than IEMOCAP.
Even though utterances in Crema-D are shorter than IEMOCAP (see Fig.~\ref{fig:duration_hist}(c)), EERs are better for the former, which could be because phonetic content variability limited to only 12 sentences.
For comparison, the authors in~\cite{kanagasundaram2019study} report EER increasing from 2.5\% to more than 20\% when going from full-length recordings to 5sec versus 5sec trials on NIST 2010 corpora.

Also, it should be noted that EERs are worse when the test utterance emotion is different from enroll utterance emotion, suggesting that speaker verification systems are sensitive to change of emotion.
It could be a very serious problem in real scenarios because humans can easily change their emotions according to the situation.
In same-emotion trials, Neutral vs. Neutral performed best on IEMOCAP and MSP-Podcast while the same pair performed worst in Crema-D. Sad vs. Sad is best on Crema-D.
Angry vs. Angry on IEMOCAP and Happy vs. Happy on MSP-Podcast are the worst same-emotion trials.
In cross-emotion trials, Angry vs. Sad/Neutral is the worst performing emotion pair on IEMOCAP, Angry vs. Happy on MSP-Podcast and Angry vs. Sad on Crema-D.
It can be observed that emotion Angry is common in worst performing cross-emotion trials across datasets.
Except on MSP-Podcast, all cross-emotion trails performed worst compared to same-emotion trails.

\begin{table}
    \centering
    \begin{tabular}{@{}c|c|c|c|c@{}}
    \toprule
\diagbox{Enroll}{Test}& Angry & Happy & Sad & Neutral \\
\midrule
Angry &42.19 & 44.11 & 44.35 &  44.35 \\
Happy & 44.11  & 41.47  & 42.52  & 43.2 \\
Sad & 44.35  & 42.52  & 40.45  & 43.27 \\
Neutral & 44.35  & 43.2 &  43.27 &  39.4 \\
\bottomrule
    \end{tabular}
    \caption{EER for Speaker Verification on IEMOCAP}\label{tab:eer_iemocap}
\end{table}

\begin{table}
    \centering
    \begin{tabular}{@{}c|c|c|c|c@{}}
    \toprule
 \diagbox{Enroll}{Test}& Angry & Happy & Sad & Neutral \\
\midrule
Angry & 13.14  & 18.15 &  17.28 &  12.98 \\
Happy &  18.15  & 15.41 &  13.97  & 11.63 \\
Sad &  17.28  & 13.97 &  13.34  & 11.89  \\
Neutral &  12.98  &  11.63 &  11.89  & 8.95  \\
\bottomrule
    \end{tabular}
\caption{EER for Speaker Verification on MSP-Podcast}\label{tab:eer_msp}
\end{table}

\begin{table}
    \centering
    \begin{tabular}{@{}c|c|c|c|c@{}}
    \toprule
\diagbox{Enroll}{Test}& Angry & Happy & Sad & Neutral \\
\midrule
Angry & 23.6 & 30.29 &  34.65 &  32.36\\
Happy & 30.29  & 25.81 &  34.07  & 31.08 \\
Sad &  34.65 &  34.07 &  20.26 &  28.43\\
Neutral &  32.36  & 31.08 &  28.43 &  26.92 \\
\bottomrule
    \end{tabular}
\caption{EER for Speaker Verification on Crema-D}\label{tab:eer_cremad}
\end{table}

\section{Conclusions and Future Work}
\label{sec:conclusion}

In this work, we study the connections between speaker recognition and emotion recognition. 
We first show that emotion recognition performance can be improved using speaker recognition models such as x-vectors through transfer learning.
Then, we show the effect of emotion on speaker verification performance.
For emotion recognition, we observed that features extracted from pre-trained models performed better than the features curated for emotion recognition tasks such as GeMAPS.
We noticed that the unaugmented x-vector model features perform better than the augmented x-vector model features for emotion recognition.
Best emotion recognition performance on all 3 datasets is obtained by fine-tuning the pre-trained x-vector models.
Data augmentation for emotion classification provided consistent improvements on 2 out of 3 datasets. 
We observed that the unaugmented x-vector model is not robust to noise.
In terms of absolute improvement, we obtained 30.40\%, 7.99\%, and 8.61\% on IEMOCAP, MSP-Podcast, and Crema-D, respectively, over the baseline model with no pre-training.

Finally, analysis of the effect of emotion on speaker verification models revealed that the latter is highly sensitive to change in the emotion of test speakers.
We observed that same-emotion trials perform better than cross-emotion trials.
Among worst-performing cross-emotion trials, angry was common across all datasets.
As part of future work, we will focus on emotion-invariant speaker verification models.
We hope our work will initiate a new line of research in the speaker recognition community.

\bibliographystyle{IEEEbib}
\bibliography{strings,refs}

\end{document}